\documentclass[11pt,a4paper]{article}
\usepackage{graphicx}
\date{}
\usepackage[latin1]{inputenc}
\usepackage{amsmath}
\usepackage{amsfonts}
\usepackage{amssymb}
\author{R. L.  Silva, A.R. Pereira and W.A. Moura-Melo
\\Departamento de F\'isica,
Universidade Federal de Vi\c cosa, Vi\c cosa,  36570-000\\Minas
Gerais, Brazil.}
\title{{Magnetization reversals in a disk-shaped small magnet with an interface}}
\begin{document}
\maketitle
\begin{center}
\textbf{Abstract}
\end{center}

\indent We consider a nanodisk possessing two coupled materials
with different ferromagnetic exchange constant. The common border
line of the two media passes at the disk center dividing the
system exactly in two similar semidisks. The vortex core motion
crossing the interface is investigated with a simple description
based on a two-dimensional model which mimics a very thin real
material with such a line defect. The main result of this study is
that, depending on the magnetic coupling which connects the media,
the vortex core can be dramatically and repeatedly flipped from up
to down and vice versa by the interface. This phenomenon
produces burstlike emission of spin waves each time the
switching process takes place.    
Corresponding author: A. R. Pereira\\
e-mail: apereira@ufv.br\\
Tel.: +55-31-3899-2988.\\
Fax: +55-31-3899-2483.\\

\maketitle

\section{Introduction}

 \indent The study of topological excitations is an important topic in
 modern theoretical and experimental physics. These objects are
 also related to technological applications in several branches of
 condensed matter physics such as superconductivity,
 superfluidity, magnetism, etc. For example, in micrometer-sized magnetic
 thin  films, the magnetization typically adopts an in-plane circular configuration
 known as a magnetic vortex. At the vortex core, the magnetization
 turns sharply out of plane, pointing either up or down. Such a binary phenomenon
 generates the possibility of developing magnetic data storage but it would require
 the ability to flip the vortex core on demand. However, vortices are highly
 stable and therefore, very strong magnetic fields were previously thought
 to be necessary to accomplish this. Recently, it was shown that elaborated
 experiments with low fields (about 1.5 mT) can also reverse the
 direction of a vortex core ~\cite{Waeyenberge06}.

On the other hand, the vortex-defect interaction is another
mechanism with potential technological relevance. Really, point
defects (holes) have been intentionally incorporated in magnetic
nanodisks ~\cite{Rahm03,Rahm04,Rahm+04}. In these circumstances,
the vortex-hole interactions lead to interesting effects
~\cite{Rahm03,Rahm04,Pereira03,Pereira05,Compton06,Kuepper07,Arpereira05,PereiraJAP,ARPJAP07}.
Here, we would like to take into account another possible type of
defect in the disk, which is more associated to a line defect. We
will illustrate this in the case of a circular magnetic film
having two media with different ferromagnetic exchange coupling
constants $A_{\alpha}$ and $A_{\beta}$. Of course, these media are
separated by an one-dimensional interface and the linking between
them is achieved by a coupling constant $A_{\alpha\beta}$. A
natural question to ask is what happens to a magnetic vortex when
its core encounters such inhomogeneities in the film. It is
interesting because in thin or ultrathin ferromagnetic films, a
large fraction of the magnetic moment bearing ions sit in
interface or surface sites. These also can be affected by the
chemical absorption of selected molecules. Hence we consider an
interface (a line) passing exactly at the disk center and
therefore,  dividing the system in exactly two alike
semidisks. Note that, in principle, such an arrangement could be
built by joining two micrometer-size semidisks made of different
ferromagnetic materials. Also, these different materials (with
coupling constants $A_{\alpha}$ and $A_{\beta}$) could be suitably
chosen as required by experimental observations and available
systems. Such possibilities follow from the fact that the strength
and character of magnetic anisotropies are subject to design (spin
engineering). We would therefore expect that quasi-two-dimensional
realizations of the above theoretical system can be made for
experiments.

\section{The model and results}

\indent To justify our model we remember that in ferromagnetic
nanodisks or thin films, magnetostatic interactions usually induce
the magnetization to lie parallel to their surfaces. Therefore,
the magnetic moments will form rotationally symmetric patterns
that follow closed flux lines. At the center, the tightly wound
magnetization can not lie flat, because the short-range exchange
interaction favors a parallel alignment of neighboring spins
(magnetic moments). The direction of the out-of-plane spin
component (up or down) defines the vortex core polarization. This
configuration is the ground state and is known as vortex state. To
reproduce these properties we consider a two-dimensional square
lattice inside a circumference of radius $R$ (in the xy-plane)
possessing an ``easy-plane anisotropy" on its surface (top or
bottom). In addition, there is another kind of anisotropy on its
circumferential border that tends to make the spins to point out
along the tangent of the circumference envelop. These anisotropies
imitate the magnetostatic effects. Thus, with the above
considerations, a disk with an interface would be described by the
following Hamiltonian
\begin{eqnarray}\label{Exchange-magnetostatic-interface}
H &=& -\sum_{r=\alpha,\beta}A_{r} \sum_{\{i,j\}\in r}
\vec{\mu}_{i,r}\cdot \vec{\mu}_{j,r}- A_{\alpha\beta} \sum_{k}
\vec{\mu}_{k,\alpha}\cdot \vec{\mu}_{k+1,\alpha} \nonumber\\&& +
\sum_{\nu=1,2} \sum_{l}\delta_{\nu}(\vec
\mu_{l}\cdot\hat{n}_{l,\nu})^{2}
-\sum_{i}\vec{h}\cdot\vec{\mu}_{i},
\end{eqnarray}\\
where $\vec{\mu}_{i}=\vec{M}_{i}(\vec{r})/M_{s}=\mu_{i}^{x}\hat{x}
+\mu_{i}^{y}\hat{y}+\mu_{i}^{z}\hat{z}$ is the atomic moment
(spin) unit vector at position $i$ ($M_{s}$ is the saturation
magnetization) and $r=\alpha,\beta$ indicates the two media. In
addition, $A_{\alpha}>0$ and $A_{\beta}>0$ are ferromagnetic
couplings and the sum $\{i,j\}\in \alpha$ is over nearest-neighbor
spins of medium $\alpha$ while the sum $\{i,j\}\in \beta$
considers only nearest-neighbor spins of medium $\beta$. All spins
inside the film have coordination number of four but spins of both
edge of the interface, interact only with three other spins of the
same nature; the remaining interaction is with its
nearest-neighbor of the other medium.  This fact is included in
the second term of Hamiltonian
(\ref{Exchange-magnetostatic-interface}): the coupling between
nearest-neighbor atoms belonging to different media is given by
$A_{\alpha\beta}>0$ and therefore, $k$ indexes only magnetic
moments belonging to the line interface at the side of medium
$\alpha$ (hence, $k+1$ indexes only spins along the line interface
at medium $\beta$). The third term mimics the magnetostatic
energies and we will assume here that it does not depend on the
media. Finally, the last term of the Hamiltonian considers the
effects of an external magnetic field $\vec{h}$.

The third term ("magnetostatic energy") of Hamiltonian
~(\ref{Exchange-magnetostatic-interface}) is justified as follows:
the sum over sites $\{l\}$ considers the scalar product of the
local magnetic moments and the unitary vectors $\hat{n}_{l,\nu}$,
which are perpendicular to either the circle plane ($\nu=1$) or
the circumference contour of the film ($\nu=2$). In the present
model, this contour is the lateral border of the disk. Therefore,
the sum on $\nu$ forces the spins to preferentially become
parallel to the film surface and circumferential contour line. For
usual films (without defects, i.e.,
$A_{\alpha}=A_{\beta}=A_{\alpha\beta}$), the use of adequate
values of the parameters can make the above model to qualitatively
reproduce the vortex ground state and also, as will be seen below,
it brings about the vortex core dynamics already obtained
analytically \cite{Guslienko02} and observed in experiments
\cite{Park03,Choe04,Novosad05} as well as in micromagnetic
simulations \cite{Guslienko02,Park03,Choe04}. In principle, such a
system could represent a very thin disk with thickness $L$ and
radius $R$ so that its aspect ratio $L/R \ll 1$.

As it is well known, for usual films, the vortex structure can be
set into a gyrotropic motion by application of a small magnetic
field. This in-plane gyrotropic motion is the lowest excitation
mode in elements exhibiting a vortex structure. The sense of
gyration of the core in a circular trajectory (clockwise or
counterclockwise) is determined only by the vortex core
polarization. Because the sense of gyration determines the
polarization, it is clear that a change in it unambiguously
indicates a change in the vortex core polarization. Therefore, if
the vortex core motion could be reflected in someway, its sense of
gyration would be drastically changed and the polarization would
be reversed. Here we show that such a mechanism could be naturally
triggered by the existence of two different media separated by an
interface inside the disk. Our study considers several values of
the ratio $A_{\beta}/A_{\alpha}=\epsilon \leq 1$ with the
parameter $A_{\alpha\beta}$ ranging from $0$ to $A_{\alpha}$.

The results are obtained by using spin dynamic simulations for
spins occupying all possible points of a square lattice inside the
circumference of radius $R$ (in most simulations it was used
$R=20a, 25a$ and $30a$, where $a$ is the lattice spacing). The
Heisenberg equation of motion
$d\vec{\mu}_{i}/dt=i[\vec{\mu}_{i},H]$ is solved for each spin
$\vec{\mu}_{i}$ interacting with its nearest neighbors. We have
employed the fourth-order predictor-corrector method. In order to
excite the lowest excitation mode (gyrotropic mode), a sinusoidal
external magnetic field $\vec{h}=h_{0}\hat{x}\sin(\omega t)$ is
applied for a short time (of the order of $700A_{\alpha}^{-1}$,
which can be compared with the total time of the simulations
$10^{4}A_{\alpha}^{-1}$). The time increment is equal to $\Delta
t=0.0001A_{\alpha}^{-1}$. During all observations, the energy is
conserved and the spin constraint $\vec{\mu}^{2}=1$ remains
unaffected. In all calculations presented here the
``magnetostatic" parameters are $\delta_{1}=0.2A_{\alpha}$ and
$\delta_{2}=2A_{\alpha}$ (the disk size is $R=20a$). The essential
physics is not altered by other choices of the values of
$\delta_{1}$ if $0<\delta_{1}<0.28$. For $\delta_{1}>0.28$, the
vortex becomes essentially planar and does not develop the
out-of-plane component at the core ~\cite{Wysin94}. It is
important to say that these selected relevant parameters for the
disk ($\delta_{1}/A_{\alpha}$, $\delta_{2}/A_{\alpha}$) are
reasonable for a material such as Permalloy (Py). To see this we
carry out an estimate for a typical nanomagnet without defects
($A_{\alpha}=A_{\beta}=A_{\alpha\beta}$): in model
(\ref{Exchange-magnetostatic-interface}), the exchange length
$l_{0}$ can be written as $l_{0}\approx
a\sqrt{A_{\alpha}/\delta_{1}}$. For comparison with experimental
results, $A_{\alpha}\equiv AL$, where $A\approx 1.3\times 10^{-11}
J/m$ is the exchange constant and the thickness $L$ is on the
order of $10^{-9}m$ for thin nanodisks made of Py, while
$l_{0}\approx \sqrt{A/\mu_{0}M_{s}^{2}}$, with $M_{s}=8.6\times
10^{5}A/m$ and $\mu_{0}=4\pi \times 10^{-7}N/A^{2}$. Therefore, we
expect that $a\sqrt{A_{\alpha}/\delta_{1}} \sim
\sqrt{A/\mu_{0}M_{s}^{2}} $, which leads to
$\delta_{1}/A_{\alpha}\approx \mu_{0}M_{s}^{2}a^{2}/A \sim
10^{-1}$. It is on the same order of the values used here. On the
other hand, the range of $\delta_{2}/A_{\alpha}$ was chosen to
adequately obtain results about the vortex core dynamics
comparable with the low-frequency gyrotropic mode, which lies in
the GHz range. Really, combining the values
$\delta_{1}/A_{\alpha}$ and $\delta_{2}/A_{\alpha}$, we get a GHz
frequency $\omega_{G}\sim 0.0056A_{\alpha}$ for the circular
motion (gyrotropic mode) of the vortex core. Hence, for a disk
with the above parameters and without defects, our results are in
qualitatively and reasonably quantitatively agreement with
experimental observations, particularly for $R=20a \approx 20l_{0}
\approx 10^{2} nm$.

\begin{figure}
\begin{center}\resizebox{7cm}{!}{\includegraphics{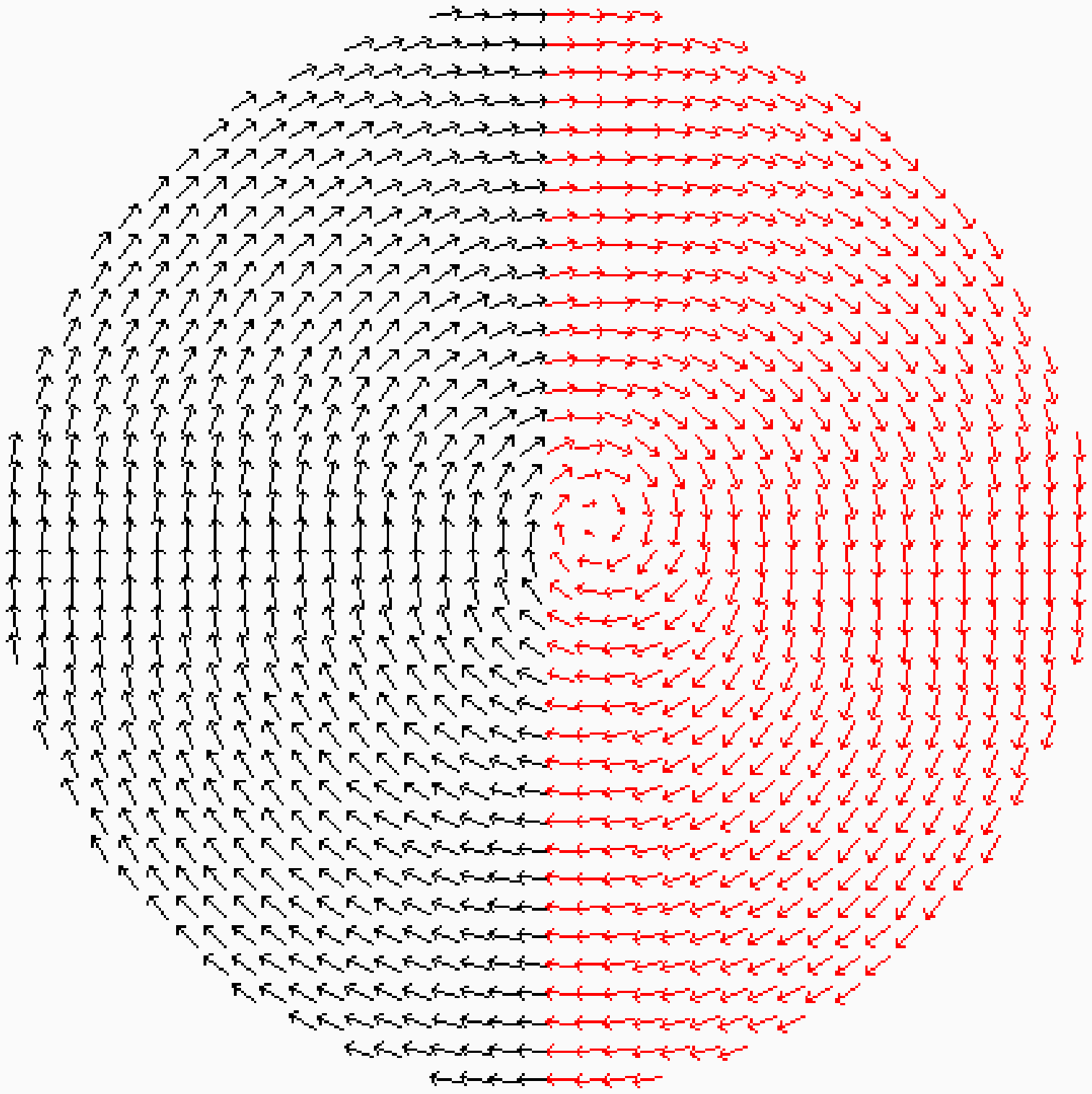}}\end{center}
\caption{} \label{2d}
\end{figure}

\begin{figure}
\begin{center}\resizebox{7cm}{!}{\includegraphics{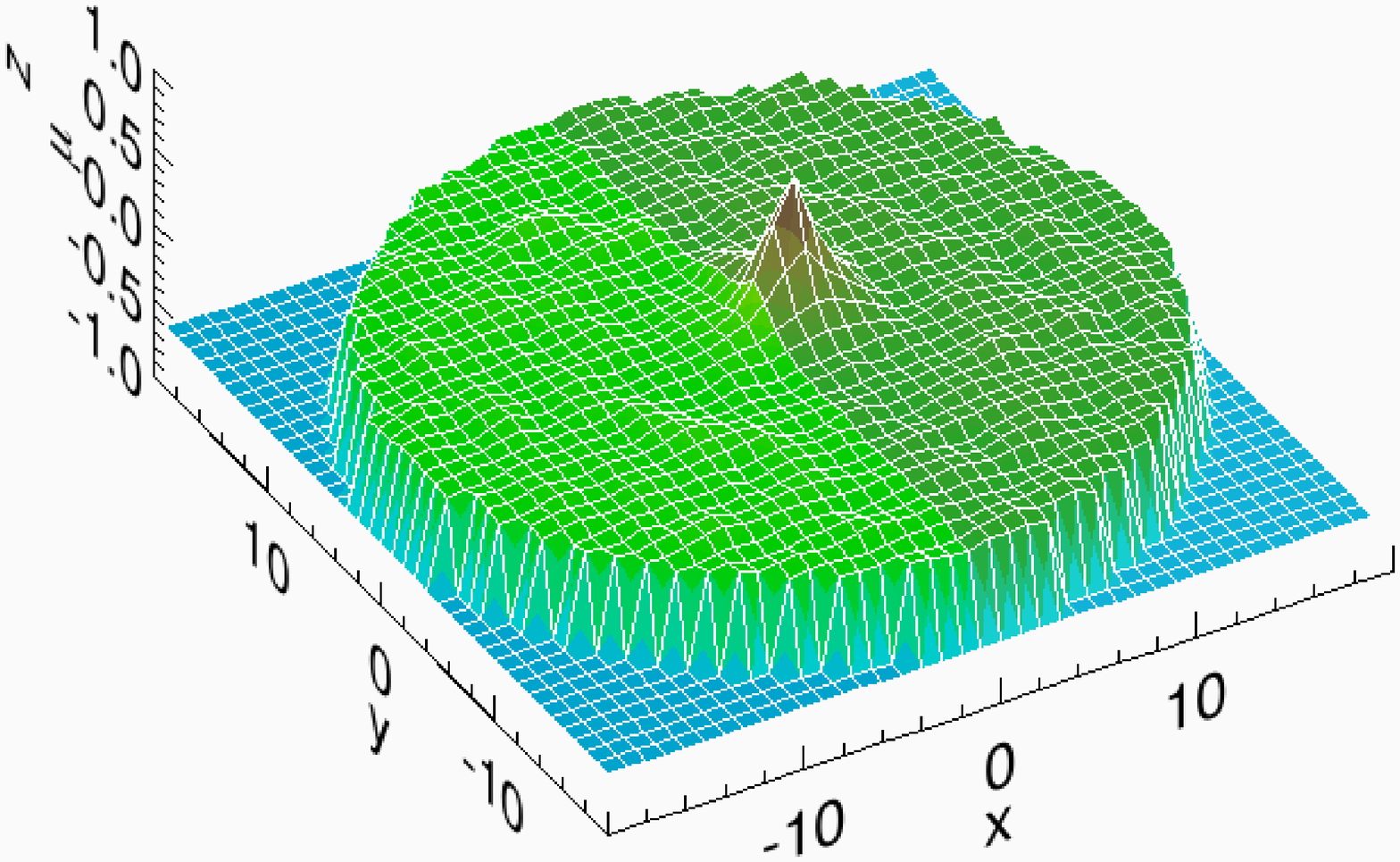}}\end{center}
\caption{}\label{3d}
\end{figure}

Of course, we expect that a film containing two different coupled
materials will exhibit distinct properties (see Figs.\ref{2d} and
\ref{3d} for visualization of a vortex in the system studied
here). Our aim now is to know how a confined vortex experiences a
line defect for some values of the parameters. Two situations
resume the main possibilities. We start studying the case for
$\epsilon=0.9$, choosing $A_{\alpha\beta}=0.8A_{\alpha}$ for the
coupling between the media. The ground state is a vortex
centralized out the disk center. Indeed, the equilibrium position
was obtained using Monte Carlo calculations at low temperatures
and, in this case, it is slightly displaced to the medium with
smaller exchange constant (medium $\beta$) at position $\approx
(a,0)$. In general, this displacement increases as $\epsilon$
decreases. The sinusoidal field used to excite the gyrotropic mode
has amplitude $0.0085A_{\alpha}$ and frequency $0.0089A_{\alpha}$.
The results show clearly that the vortex core interacts with the
interface and the gyrotropic mode becomes centralized in the
medium with smaller exchange. In addition, when the core crosses
the interface towards medium $\beta$, it speeds up (with respect
to its velocity in the medium $\alpha$) and when it goes from
medium $\beta$ to $\alpha$ the motion slows down. We also plot the
average magnetization in the $x$ and $z$ directions $<\mu^{x}>$
and $<\mu^{z}>$ in Figs. \ref{mx08} and \ref{mz08} respectively.
We notice that $<\mu^{x}>$ oscillates around zero indicating an
almost circular motion and that the frequency of this ``deformed
gyrotropic mode" is $0.00377 A_{\alpha}$, which is smaller than
$\omega_{G}$ for a disk with the same size without defects. One
may think in the possibility to use such ``exotic" systems for the
control of gyrotropic frequencies, which may be useful in
technological applications. The analysis of the out-of-plane
fluctuations is shown in Fig. \ref{mz08}. Note that $<\mu^{z}>$
oscillates very rapidly and with small amplitude around a small
but finite positive value, indicating that the core is pointing
up. These very small oscillations are essentially localized in the
vortex core as it can be seen in Fig. \ref{mzcore08}, which shows
the oscillations in $\mu^{z}$ averaged only over the core, rather
than the whole disk. In this case, the amplitude is much larger
than the one in Fig. \ref{mz08} because much less spins are
considered in the average. Such oscillations are induced by the
discreteness of the lattice of the model adopted here. In fact, in
its travel, the geometrical center of the vortex core alternates
around points containing sites and vacancies: when it becomes
centered on a site of the lattice, the average $<\mu^{z}>_{core}$
is larger due to the presence of a central spin in the core. On
the other hand, when it moves away from the site (for example, in
direction to the middle of a plaquette), the average decreases
because there is no spin in the core center. We notice that
$<\mu^{z}>_{core}$ oscillates around a value near 0.95. Since the
small oscillation in $<\mu^{z}>$ is associated with discreteness
effects, its Fourier transform does not reflect the main
characteristics of the core motion. Indeed, there are two main
peaks in its Fourier transform, which are not very different from
that of the case free of defects. Figure \ref{orbitaface}
summarizes the main properties of the vortex motion showing the
trajectory followed by the core during several laps.

No switching process is observed for these parameters. On the
other hand, for $A_{\alpha\beta}=0.5$, interesting phenomena take
place. The vortex core equilibrium position is $\approx(a,0)$ (see
Figs.~\ref{2d} and ~\ref{3d}) for the $2D$ and $3D$ views of the
system respectively) and it becomes completely confined in the
medium $\beta$, even after starting its motion (induced by the
sinusoidal field with amplitude $0.007A_{\alpha}$ and frequency
$0.089A_{\alpha}$). Indeed, the core moves initially along a
straight line (almost perpendicular to the interface) until a
distance $d<R$ from the disk center and then it goes back to the
interface not by the same path but following an approximate
circular trajectory. However, arriving at the interface, the core
is reflected by the line defect and consequently, a magnetization
reversal takes place, causing a burstlike emission of spin waves.
In sequence, the core goes back through almost the same circular
trajectory until finding the interface again at diametrally
opposite point, where it is reflected (and flipped) for a second
time and so on (see Fig.~\ref{semiorbita} and supplementary Video
of the vortex core dynamics as auxiliary material \cite{VIDEO}).
Therefore, only a semi-circular mode is observed. Indeed, the core
tries to develop a complete gyrotropic motion (see also
Figs.~\ref{mx} and ~\ref{my}), but it is impeded by the interface.
The Fourier transform of $<\mu^{x}>$ has a main peak at frequency
$0.0039 A_{\alpha}$, while for $<\mu^{y}>$ the main peak occurs at
frequency $0.0081A_{\alpha}$. Hence, the motion becomes very
irregular. The amplitude of the vortex motion decreases as this
process is repeated and the vortex center speeds up leading to
rapid changes in the directions of motion and the subsequent
phenomenon in which the core magnetization changes coherently up
and down. The core approaches more and more the defect and
eventually it becomes trapped, oscillating along the straight line
of the interface alike an one-dimensional damped harmonic motion
with frequency $0.027A_{\alpha}$. Of course, this additional frequency is
present only in the Fourier transform of $<\mu^{x}>$, since the
vortex center now oscillates only along the interface (y-axis).
All these up-down, down-up sequences are really intense occurring
surrounded by a ``sea" of spin waves, which are all the time
perturbing the vortex core. These dramatic and sudden switching
processes are caused by the competition between magnetostatic
interactions and the strong discontinuity in the exchange
interaction along the interface. These are truly exchange
explosions and spin wave fluctuations are strongly produced. Such
phenomenon may provide further insights for generating and controlling
spin waves in magnetic nanodisks.

In order to see more details about this switching mechanism, we
have also plotted the average magnetization along the z-direction
$<\mu^{z}>$ in Fig.~\ref{mz}. It can be easily observed that,
initially, $<\mu^{z}>$ alternates its rapid oscillations between
very small positive and negative values. As already discussed
earlier, the mechanism of this effect is the sequential
magnetization reversals of the vortex core due its interactions
with the line defect. This effect remains until $t\sim
5500A_{\alpha}^{-1}$. After that, the out-of-plane magnetization
oscillates still more rapid around zero. This characterizes the
capture process of the core by the interface; although the vortex
center oscillates along the line defect, its out-of-plane core is
almost lost. Really, it seems that after being captured by the
interface, the vortex center still remains oscillating with some
reversal.

\section{Conclusions and prospects}

\indent Magnetization switching is a remarkable effect observed in
a broad range of magnetic materials. Here, we have presented a
study of how a defect along a line separating two magnetic media
may induce polarization reversal in vortex-like magnetization
lying on nanomagnets. Using Hamiltonian
~(\ref{Exchange-magnetostatic-interface}) we have studied the
vortex core interactions with the line defect in confined magnetic
systems. In the absence of defects, the results obtained by this
model agree qualitatively with experimental observations.
Therefore, it is sound natural to expect that we can be able to
predict still unobserved facts and possibilities for disks possessing defects
and other characteristics.

As discussed before, a change in the sense of gyration of the
vortex structure is an unambiguous indication of a switching of
the vortex core polarization. So, we have predicted a sequential
switching process induced by vortex-interface interactions.
Indeed, for $\epsilon=0.9$ and $A_{\alpha\beta}$ not so large, we
easily observe that every time the core is reflected by the
interface, the magnetization reversal happens \cite{VIDEO}. It
prevails for some time interval and eventually, the core becomes
captured by the defect. After being captured, the vortex core
appears to oscillate forward-backward along the line defect,
reversing its remaining polarization (which is much smaller than
before its capture, as it can be seen in video \cite{VIDEO}.
During these oscillations, a considerable amount of spin waves is
emitted, mainly when the core reversals take place. Such bursts of
spin waves propagate throughout the system, interacting and
disturbing the magnetization dynamics as a whole. Reaching the
disk border, such waves are reflected back, so that larger and
lager amount of spin waves occupies the system. Eventually, the
vortex dynamics runs out once all its kinetic energy were
dissipated. Therefore, our original proposal of investigating the
effects of a line on vortex dynamics seems to be also very useful
to produce such waves and study their properties in confined
structures.

In summary, we have investigated the vortex core dynamics in
magnetic nanodisks with a line defect. Our model predicts
remarkable switching process which may be useful for technological
applications such as those which utilize vortex degrees of
freedom, namely, its polarization and mechanisms associated to its
reversal. Very recently, such a phenomenon has been observed
experimentally \cite{Gao2008} and in simulations
\cite{RicardoPRB2008,JAP2008Moura} in nanomagnets with artificial
holes.  As prospects for future investigations, we may quote the
use of line defects to divide a magnetic sample in several islands
separated by such interfaces. For instance, whenever one can set
up a vortex in each island, one can produce an array with possible
interesting interactions between vortices located at neighbor
islands. Perhaps, a global control of vortex properties
(polarization, chirality, etc) could be attainable by means of
such interactions.

The authors thank CNPq, FAPEMIG and CAPES (Brazilian agencies) for
financial support.

\begin{figure}
\begin{center}\resizebox{7cm}{!}{\includegraphics{mx08.eps}}\end{center}
\caption{}\label{mx08}
\end{figure}

\begin{figure}
\begin{center}\resizebox{7cm}{!}{\includegraphics{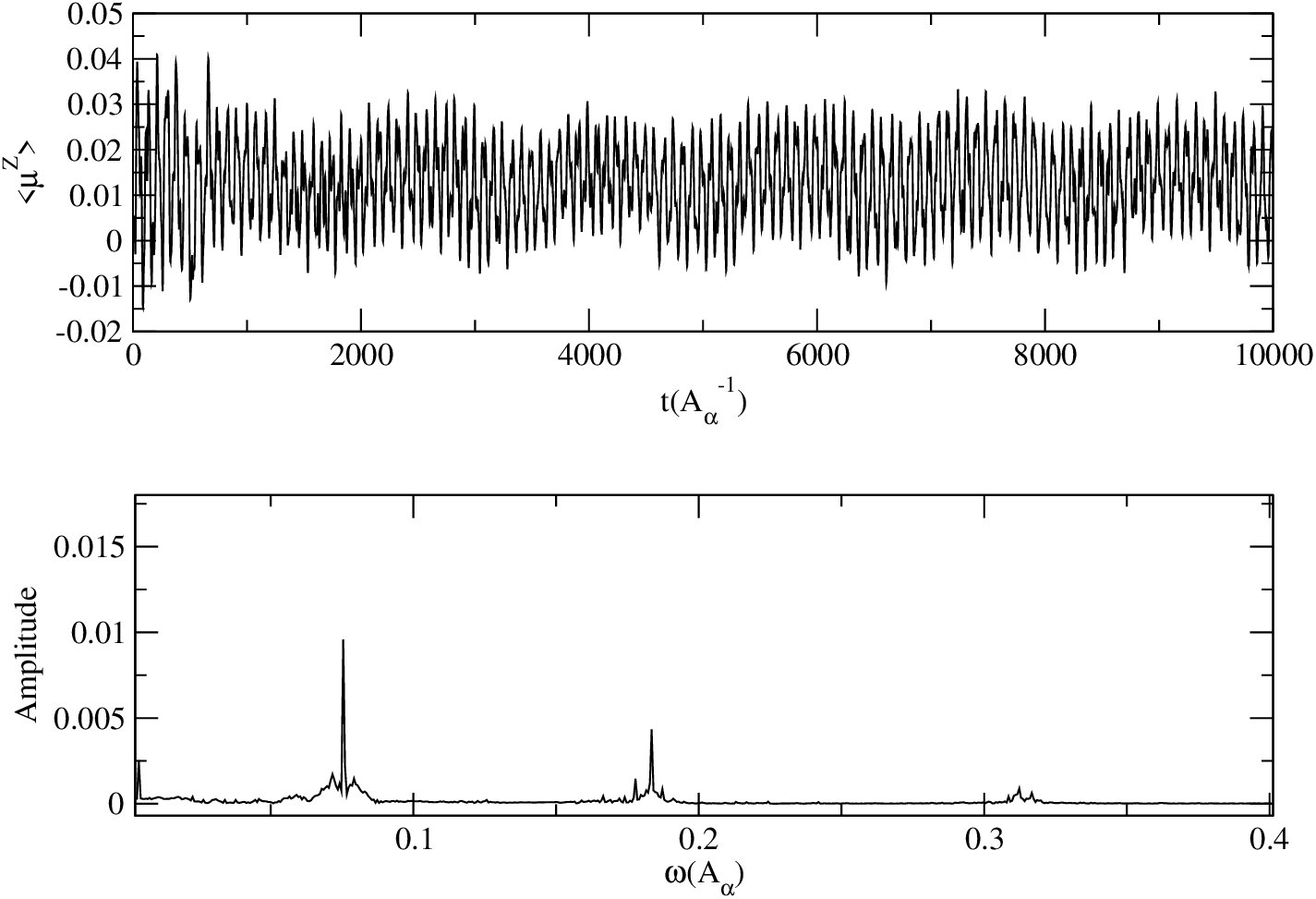}}\end{center}
\caption{}\label{mz08}
\end{figure}
\begin{figure}
\begin{center}\resizebox{7cm}{!}{\includegraphics{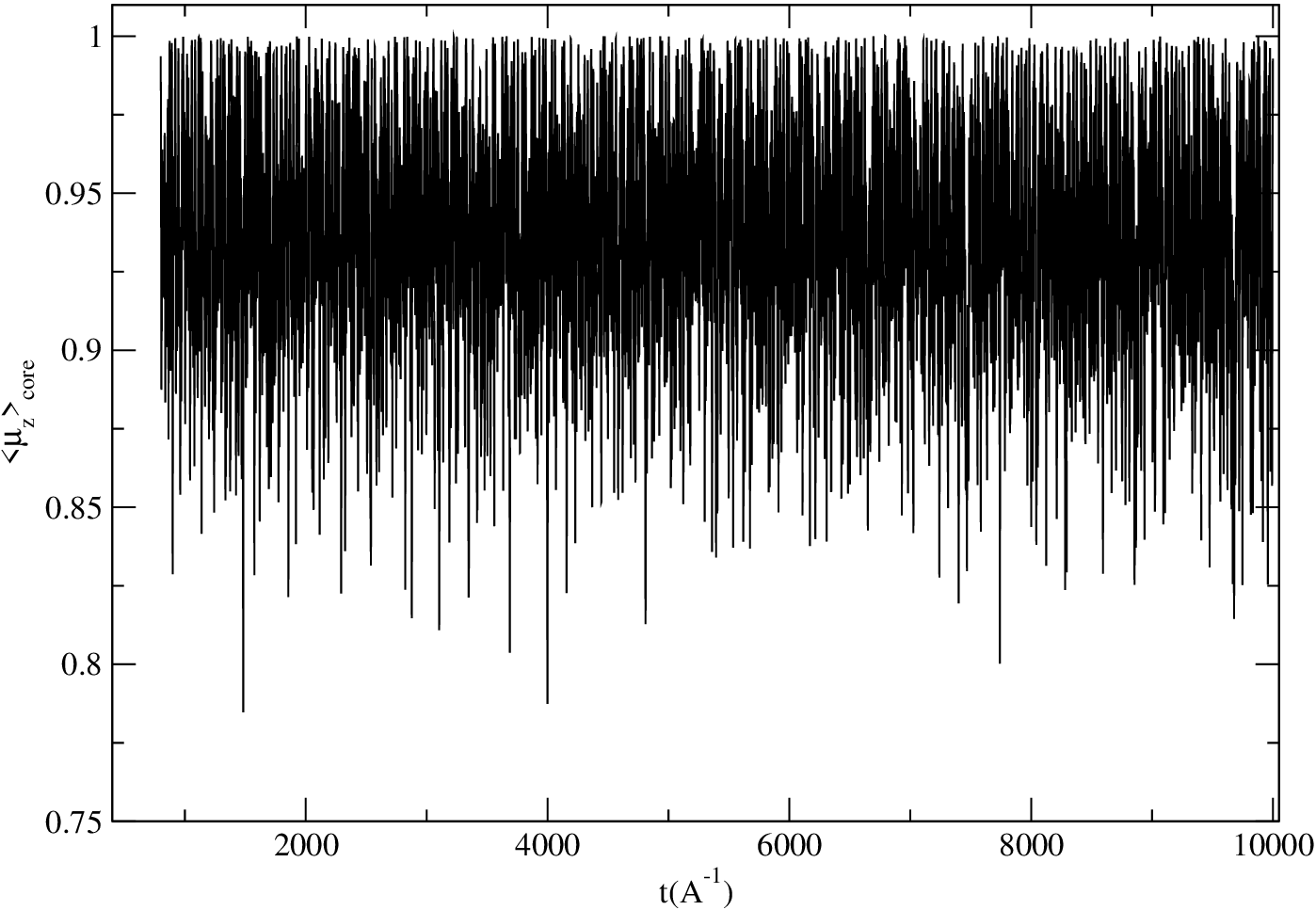}}\end{center}
\caption{}\label{mzcore08}
\end{figure}
\begin{figure}
\begin{center}\resizebox{7cm}{!}{\includegraphics{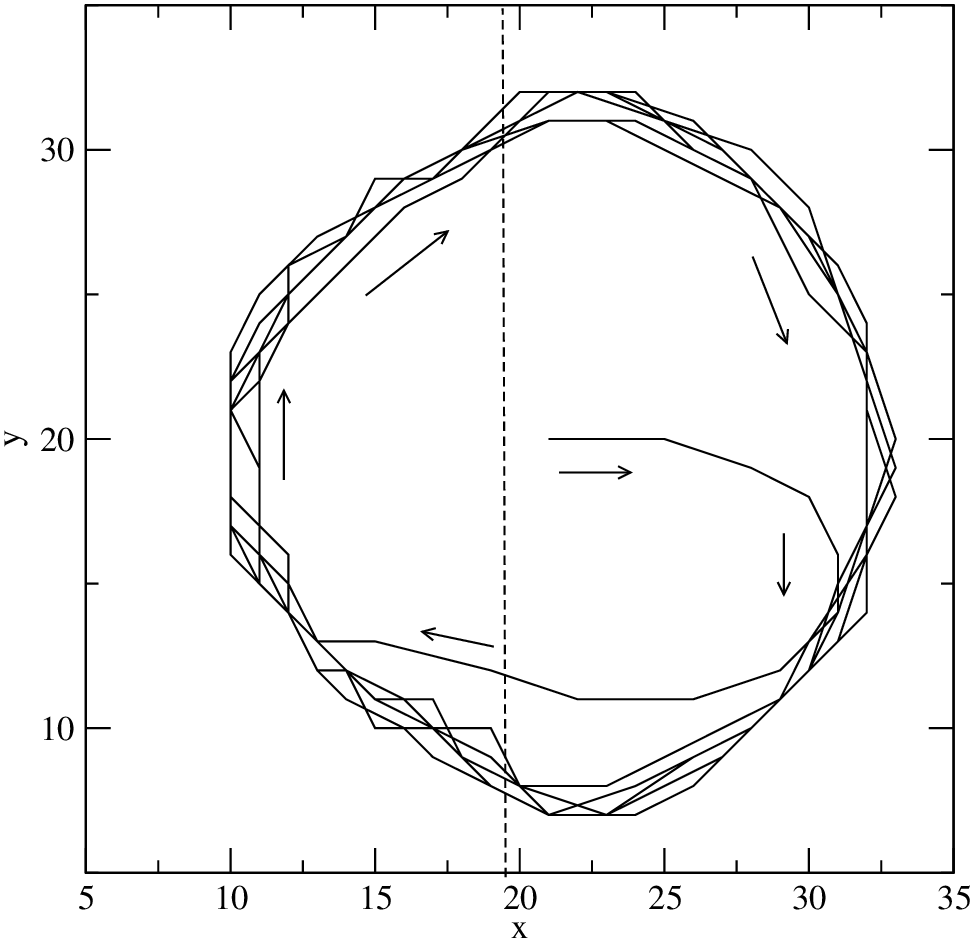}}\end{center}
\caption{}\label{orbitaface}
\end{figure}

\begin{figure}
\begin{center}\resizebox{7cm}{!}{\includegraphics{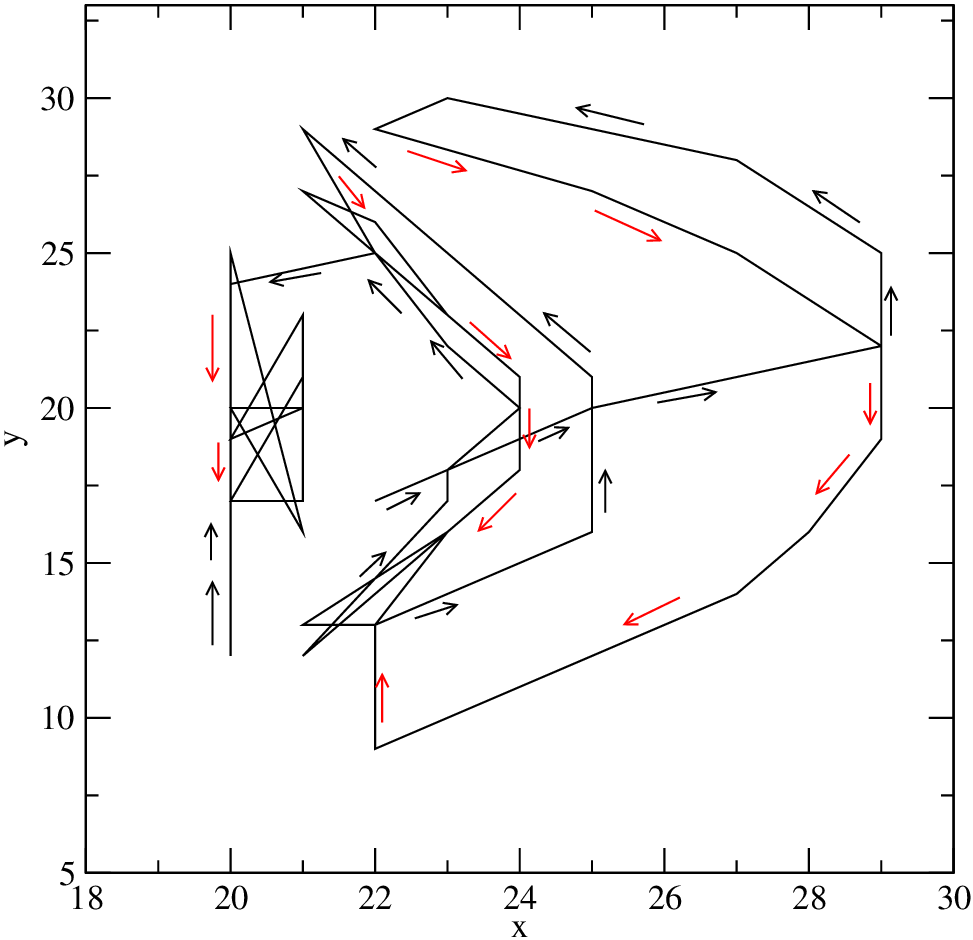}}\end{center}
\caption{}\label{semiorbita}
\end{figure}

\begin{figure}
\begin{center}\resizebox{7cm}{!}{\includegraphics{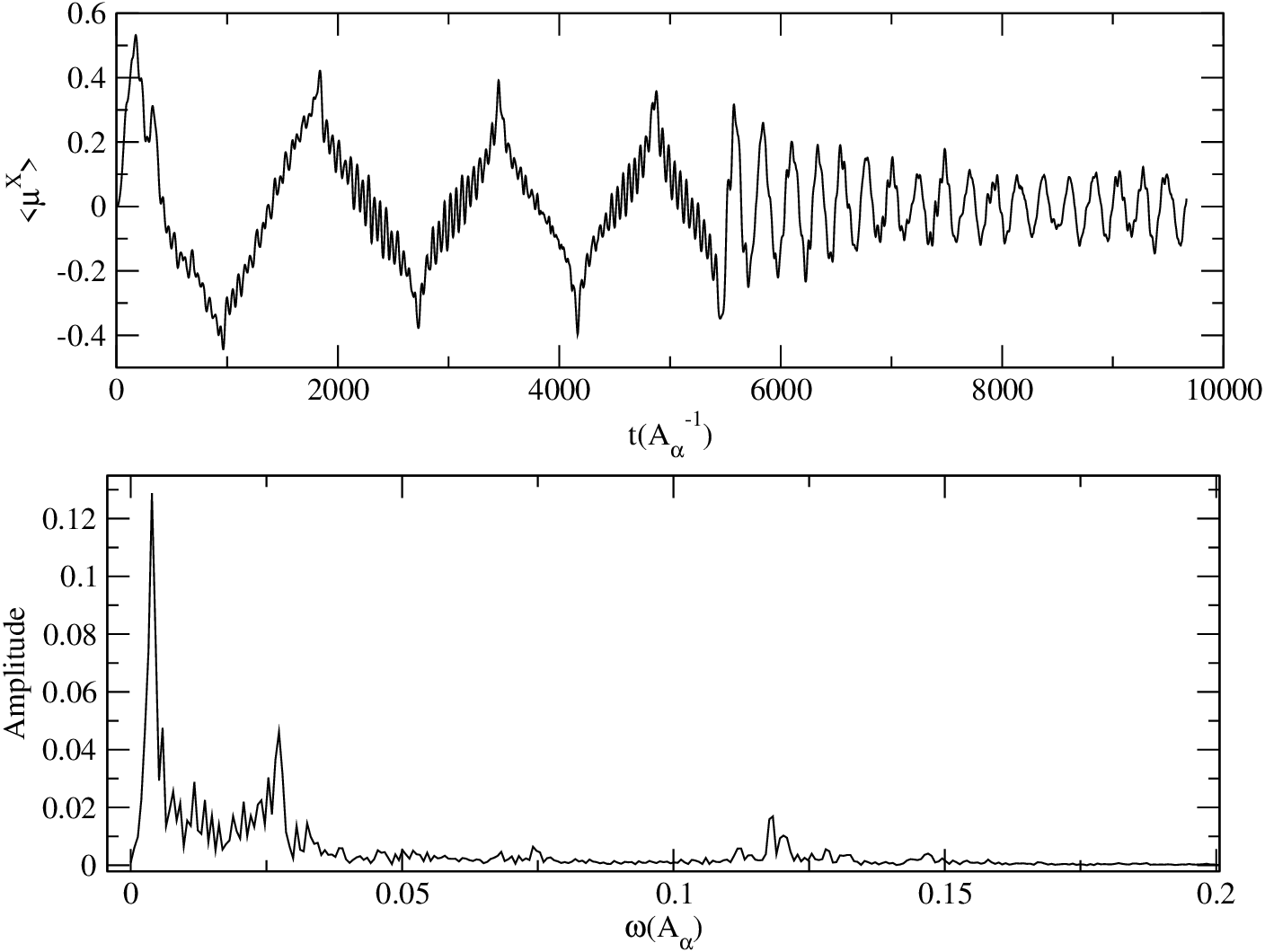}}\end{center}
\caption{}\label{mx}
\end{figure}

\begin{figure}
\begin{center}\resizebox{7cm}{!}{\includegraphics{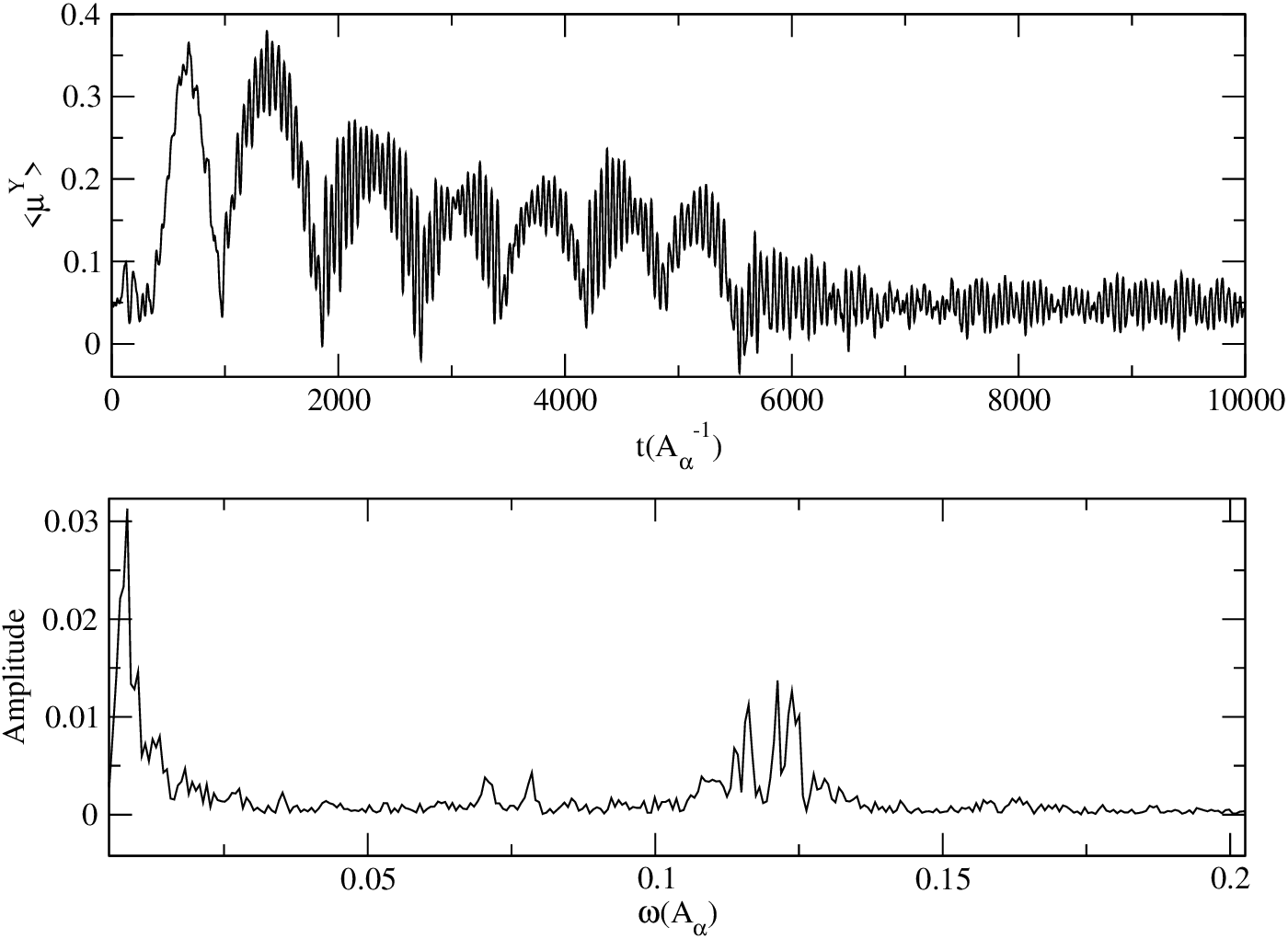}}\end{center}
\caption{}\label{my}
\end{figure}

\begin{figure}
\begin{center}\resizebox{7cm}{!}{\includegraphics{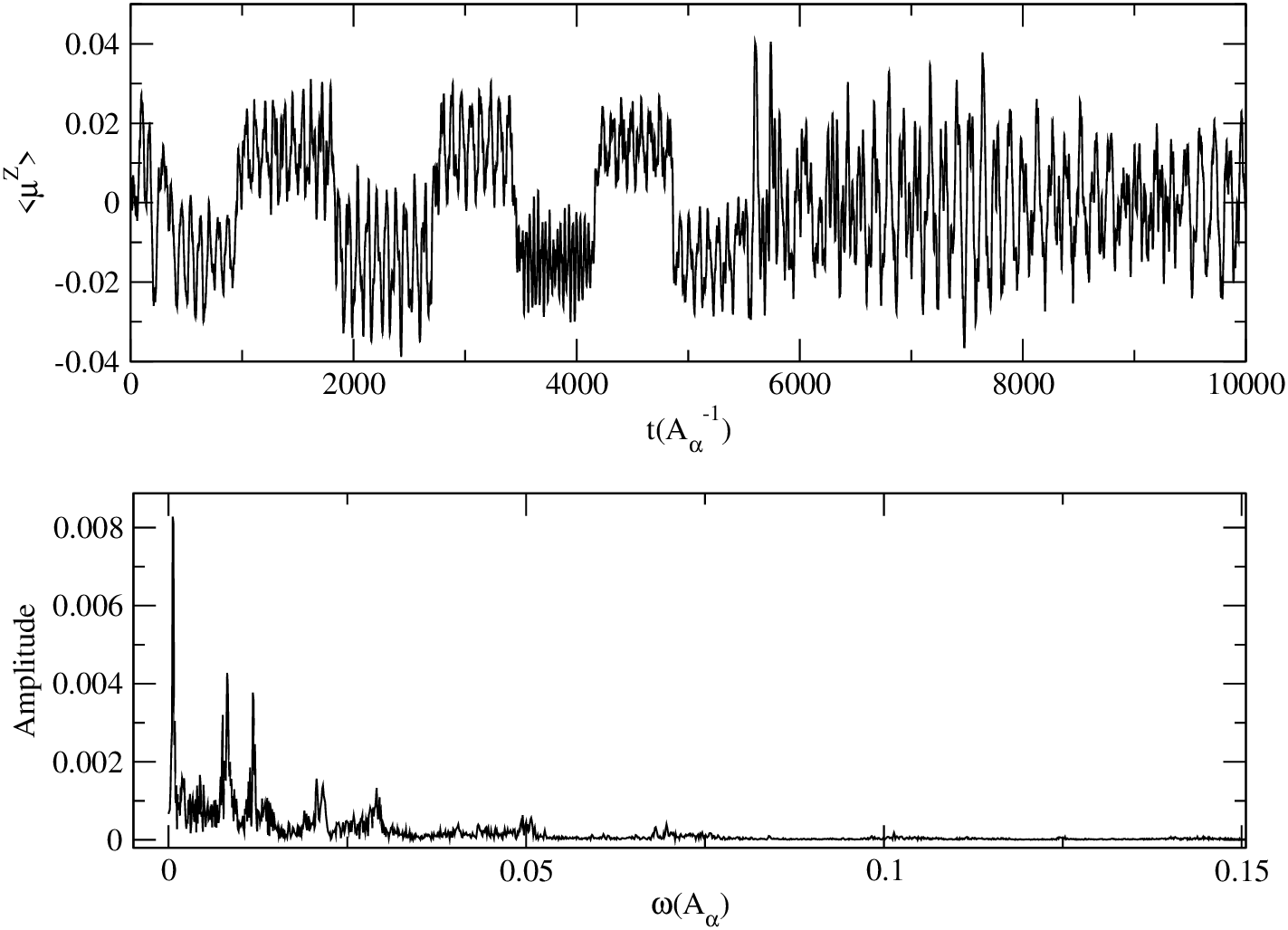}}\end{center}
\caption{}\label{mz}
\end{figure}


\newpage
\begin{thebibliography}{0}


\bibitem{Waeyenberge06} B. Van Waeyenberge, A. Puzic, H. Stoll, K.W. Chou, T. Tyliszczak,
R. Hentel, M. F\"{a}hnle, H. Br\"{u}ckl, K. Rott, G. Reiss, I.
Neudecker, D. Weiss, C.H. Back, and G. Sch\"{u}tz., Nature
\textbf{444}, 461 (2006).
\bibitem{Rahm03} M. Rahm M., J. Biberger, V. Umansky, and D. Weiss, J. Appl.
Phys. \textbf{93}, 7429, (2003).
\bibitem{Rahm04} M. Rahm, R. H\"{o}llinger, V. Umansky, and D. Weiss, J.
Appl. Phys.\textbf{95},6708 (2004).
\bibitem{Rahm+04} M. Rahm, J. Stahl, W. Wegscheider, and D. Weiss, Appl.
Phys. Lett.\textbf{85}, 1553 (2004).
\bibitem{Pereira03} A.R. Pereira, L.A.S.  M\'{o}l, S.A. Leonel, P.Z. Coura, and B.V. Costa,
Phys. Rev. B \textbf{68}, 132409, (2003).
\bibitem{Pereira05} A.R. Pereira, S.A. Leonel, P.Z. Coura, and B.V.
Costa, Phys. Rev. B \textbf{71}, 014403 (2005).
\bibitem{Compton06} R.L. Compton and P.A. Crowell, Phys. Rev. Lett.
\textbf{97}, 137202 (2006).
\bibitem{Kuepper07} K. Kuepper, L. Bischoff, Ch. Akhmadaliev, J. Fassbender, H. Stoll,
K.W. Chou, A. Puzic, K. Fauth, D. Dolgos, G. Sch\"{u}tz, B. Van
Waeyenberge, T. Tyliszczak, I. Neudecker, G. Woltersdorf, and C.H.
Back, Appl. Phys. Lett.\textbf{90}, 062506 (2007).
\bibitem{Arpereira05} A.R. Pereira, Phys. Rev. B \textbf{71},224404 (2005).
\bibitem{PereiraJAP} A.R.Pereira, J. Appl. Phys. \textbf{97}, 094303 (2005).
\bibitem{ARPJAP07} A.R. Pereira, A.R. Moura, W.A. Moura-Melo, D.F. Carneiro,
S.A. Leonel, and P.Z. Coura, J. Appl. Phys. \textbf{101}, 034310,
(2007).
\bibitem{Guslienko02} K. Yu. Guslienko, B.A. Ivanov, V.
Novosad, Y. Otani, H. Shima, and K. Fukamichi, J. Appl. Phys.
\textbf{91}, 8037 (2002).
\bibitem{Park03} J.P. Park, P. Eames, D.M. Engebretson, J. Berezovsky, and P.A. Crowell
Phys. Rev. B \textbf{67}, 020403 (2003).
\bibitem{Choe04} S.-B. Choe, Y. Acremann, A. Scholl, A. Bauer, A. Doran, J. St\"{o}hr,
and H.A. Padmore, Science \textbf{304}, 420 (2004).
\bibitem{Novosad05} V.Novosad, F.Y. Fradin, P.E. Roy, K.S. Buchanan, K. Yu. Guslienko,
and S.D. Bader, Phys. Rev. B \textbf{72}, 024455 (2005).
\bibitem{Wysin94} G.M. Wysin, Phys. Rev. B \textbf{49}, 8780 (1994).
\bibitem {VIDEO} Auxiliary video, clearly showing the vortex core motion interacting with the interface
and subsequent switching processes, is available under request:
ricardodasilva@ufv.br, apereira@ufv.br, or winder@ufv.br .
\bibitem{Gao2008} X.S. Gao, A.O. Adeyeye, and C.A. Ross, J.
Appl. Phys.\textbf{103}, 063906 (2008).
\bibitem{RicardoPRB2008} R.L. Silva, A.R. Pereira, R.C. Silva, W.A. Moura-Melo,
N.M. Oliveira-Neto, S.A. Leonel, and P.Z. Coura, Phys. Rev. B
(2008), in press.
\bibitem{JAP2008Moura} W.A. Moura-Melo, A.R.
Pereira, R.L. Silva, and N.M. Oliveira-Neto, J. Appl. Phys.
\textbf{103}, 124306 (2008).
\newpage
Figure Captions

Fig. 1 (Color online) Top view of a nanodisk with an interface
along the y-axis. The two sides of the disk are made of different
materials (here, black with exchange coupling $A_{\alpha}$ and red
with $A_{\beta}$) and are joined by an inter-side coupling
$A_{\alpha\beta}$. The vortex core equilibrium position is located
at the medium with smaller exchange constant (red medium).

Fig. 2 (Color online) Three-dimensional view of a disk with an
interface. The vortex core can be seen in its equilibrium position
pointing ``up" in the medium with smaller exchange constant
$A_{\beta}$. The application of an external field induces the
gyrotropic mode. As the core tries to cross the interface, it can
experience several effects, depending on $A_{\beta}$ and
$A_{\alpha\beta}$.

Fig. 3 The average magnetization in the $x$-direction $<\mu^{x}>$
and its Fourier transform for $\epsilon=0.9$ and
$A_{\alpha\beta}=0.8A_{\alpha}$. The same behavior is verified for
the $y$-component of the average magnetization (not shown).

Fig. 4 The average magnetization in the $z$-direction $<\mu^{z}>$
and its Fourier transform for $\epsilon=0.9$ and
$A_{\alpha\beta}=0.8A_{\alpha}$.

Fig. 5 The average magnetization in the $z$-direction over the
vortex core $<\mu^{z}>_{core}$ for $\epsilon=0.9$ and
$A_{\alpha\beta}=0.8A_{\alpha}$.

Fig. 6 The gyrotropic motion of the vortex core in a nanodisk with
an interface for $\epsilon=0.9$ and
$A_{\alpha\beta}=0.8A_{\alpha}$. The disk center is placed at
$(20a,20a)$ and the dashed line is along the defect. Like in
Fig.~\ref{2d}, media $\alpha$ and $\beta$ are the left and right
parts of the disk, respectively. Note that the path of the core is
larger in medium $\beta$.

Fig. 7 (Color online) Vortex core motion in the disk with an
interface ($\epsilon=0.9, A_{\alpha\beta}=0.5A_{\alpha}$). The
center of the disk is placed at $(20a,20a)$ and, therefore, this
figure shows only the right part of the disk. In fact, for this
case, the vortex core becomes confined in the medium $\beta$.
Black arrows indicate the counterclockwise gyration while the red
ones indicate the clockwise.

Fig. 8 The average magnetization in the $x$-direction $<\mu^{x}>$
and its Fourier transform for $\epsilon=0.9$ and
$A_{\alpha\beta}=0.5A_{\alpha}$. Initially, the vortex core moves
along trajectories that resemble semi-circumferences until being
captured by the defect.

Fig. 9 The average magnetization in the $y$-direction $<\mu^{y}>$
and its Fourier transform for $\epsilon=0.9$ and
$A_{\alpha\beta}=0.5A_{\alpha}$.

Fig. 10 The average magnetization in the $z$-direction $<\mu^{z}>$
and its Fourier transform for $\epsilon=0.9$ and
$A_{\alpha\beta}=0.5A_{\alpha}$. Note the cyclic switching
processes until the capture around $t=5500A_{\alpha}^{-1}$.

\clearpage \hspace{40cm}
\end{thebibliography}
\end{document}